\documentclass[12pt]{article} %***
\usepackage[sectionbib]{natbib}
\usepackage{array,epsfig,rotating}
\usepackage[]{hyperref}  %<----modified by Ivan
\usepackage{sectsty, secdot}
\usepackage{comment}
\usepackage{amsmath}
\usepackage{amsmath,bbm}
\usepackage{graphicx} %,psfrag,epsf}%\dot{}
\usepackage{enumerate}
\usepackage{url} % not crucial - just used below for the URL

\renewcommand{\theequation}{\thesection\arabic{equation}}
\subsectionfont{\fontsize{12}{14pt plus.8pt minus .6pt}\selectfont}
\usepackage{amsmath}
\usepackage{amssymb}
\usepackage{amsfonts}
\usepackage{multirow}
\usepackage{amsthm}
\usepackage{algorithm}
\usepackage{algorithmic}
\usepackage{authblk}

\usepackage{caption}
\usepackage{subcaption}

\theoremstyle{definition}

\newtheorem{example}{Example}

\usepackage{setspace}
\usepackage{xcolor}

\newcommand{\bm}[1]{\mbox{\boldmath$ #1 $\unboldmath}}

\title{A Decision Analysis Framework for High-fidelity and Low-fidelity Systems 
with Applications in Manufacturing Processes}
\author[1]{Fan Zhang}
\author[2]{Qiong Zhang\thanks{Corresponding author:qiongz@clemson.edu}}
\author[3,6]{Madhura Limaye}
\author[4,5]{Dhanashree Shinde}
\author[6]{Gang Li}
\author[4]{Sai Aditya Pradeep}
\author[4,5,7,8,9]{Srikanth Pilla}

\affil[1]{Department of Mathematics, Boise State University}
\affil[2]{School of Mathematical and Statistical Sciences, 
Clemson University}
\affil[3]{Manufacturing Science Division, Oak Ridge National Laboratory}
\affil[4]{Center for Composite Materials, University of Delaware}
\affil[5]{Department of Materials Science and Engineering, University of Delaware}
\affil[6]{Department of Mechanical Engineering, Clemson University}
\affil[7]{Department of Mechanical Engineering, University of Delaware}
\affil[8]{Department of Chemical and Biomolecular Engineering, University of Delaware}
\affil[9]{Department of Computer and Information Sciences, University of Delaware}

\begin{document}
\date{}
\maketitle	
%\date{}
	%%%%%%%%%%%%%%%%%%%%%%%%%%%%%%%%%%%%%%%%%%%%%%%%%%%%%%%%%%%%%%%%%%%%%%%%%%%%%%%%%%%%%%%%%%%%%%%%%%%%%%%%%%%%%%%%%%%%%%%%%%%%
%%%%%%%%%%%%%%%%%%%%%%%%%%%%%%%%%%%%%%%%%%%%%%%%%%%%%%%%%%%%%%%%%%%%%%%%%%%%%%%%%%%%%%%%%%%%%%%%%%%%%%%%%%%%%%%%%%%%%%%%%%%%
\iffalse	
	\renewcommand{\baselinestretch}{2}
	
	\markright{ \hbox{\footnotesize\rm %Statistica Sinica
			%{\footnotesize\bf 24} (201?), 000-000
		}\hfill\\[-13pt]
		\hbox{\footnotesize\rm
			%\href{http://dx.doi.org/10.5705/ss.20??.???}{doi:http://dx.doi.org/10.5705/ss.20??.???}
		}\hfill }
	
	\markboth{\hfill{\footnotesize\rm Fan Zhang, Qiong Zhang} \hfill}
	{\hfill {\footnotesize\rm Bayesian Calibration on Injection Molding Data} \hfill}
	
	\renewcommand{\thefootnote}{}
	$\ $\par
	
	%%%%%%%%%%%%%%%%%%%%%%%%%%%%%%%%%%%%%%%%%%%%%%%%%%%%%%%%%%%%%%%%%%%%%%%%%%%%%%%%%%%%%%%%%%%%%%%%%%%%%%%%%%%%%%%%%%%%%%%%%%%%
	
	\fontsize{12}{14pt plus.8pt minus .6pt}\selectfont \vspace{0.8pc}
	\centerline{\large\bf Bayesian Calibration on Injection Molding Data}
	\vspace{2pt}
	\centerline{\large\bf Bayesian Calibration on Injection Molding Data}
	\vspace{.4cm}
	\centerline{Fan Zhang$^{1}$, Qiong Zhang$^{2}$}
	\vspace{.4cm}
	\centerline{\it $^{1}$Clemson University,}
	\vspace{.55cm} \fontsize{9}{11.5pt plus.8pt minus.6pt}\selectfont
	%%%%%%%%%%%%%%%%%%%%%%%%%%%%%%%%%%%%%%%%%%%%%%%%%%%%%%%%%%%%%%%%%%%%%%%%%%%%%%%%%%%%%%%%%%%%%%%%%%%%%%%%%%%%%%%%%%%%%%%%%%%%
\fi

\newpage

\noindent \textbf{Abstract:} 
Optimizing complex manufacturing processes often involves a trade-off between data accuracy and acquisition cost. High-fidelity data are accurate but limited, while low-fidelity data are abundant but often biased. Balancing these two sources is critical for efficient manufacturing optimization. To address this challenge, we develop a decision analysis framework based on multi-fidelity Gaussian process (GP) modeling based on the Kennedy-O'Hagan (KOH) framework. We propose a systematic Bayesian calibration approach using multi-fidelity GPs that explicitly quantifies the model discrepancy, and an algorithm that combines posterior sampling of calibration parameters with predictive sampling to characterize the distribution of optimal input settings and their associated uncertainty. These components are integrated into a five-stage practical workflow for the optimization of manufacturing processes. Through an illustrative example and two real-world applications in composite cure cycle optimization and injection molding process control, we demonstrate how the framework integrates information from both high-fidelity and low-fidelity data sources to support decision-making under parameter uncertainty.


	\def\thefigure{\arabic{figure}}
	\def\thetable{\arabic{table}}
	
	\renewcommand{\theequation}{\thesection.\arabic{equation}}

	\fontsize{12}{14pt plus.8pt minus .6pt}\selectfont

\section{Introduction}
\label{sec:intro}

\subsection{Background and Motivation}

Computer simulations are indispensable in modern manufacturing, as they enable engineers to efficiently explore complex physical phenomena and find optimal process parameter settings that would be costly or infeasible to test experimentally. However, relying on a single data source often presents a dilemma: high-fidelity data, whether derived from physical experiments or complex simulations, are accurate but expensive to obtain, limiting their availability for optimization. Conversely, computationally efficient low-fidelity models are abundant often rely on simplifying assumptions that limit their accuracy. Consequently, a key challenge in manufacturing process optimization is to combine these data sources to support efficient and robust decision-making. We illustrate this challenge through two manufacturing applications: composite cure optimization and injection molding process control, where combining high-fidelity experimental measurements with low-fidelity simulation data leads to more efficient and practical decision-making.

\begin{example}\label{ex1}
Composite Cure Simulation. \\
In composite manufacturing, 
the simulation of the curing of L-shaped laminates is done through a multi-stage process.
%L-shaped laminates are cured through a multi-stage heating process to achieve the desired material properties and minimize residual stresses and deformations. 
This problem involves optimizing the cure cycle by selecting appropriate dwell times and temperatures at different stages. The goal is to find the optimal cure cycle parameters (dwell times $t_1$, $t_2$ and temperatures $T_1$, $T_2$) that minimize warpage or deformation while satisfying constraints on the Degree of Cure (DoC). In this example, 
the data from the first stage DoC simulation serves as the 
the low-fidelity data to deformation. 
The high-fidelity data come from the full stress-deformation simulation (second stage) that incorporates DoC results and captures actual deformation behavior, but is computationally more expensive.
\end{example}

\begin{example}\label{ex2}
Injection Molding Process. \\
Injection molding is a widely used manufacturing process for producing plastic parts, where molten polymer is injected into a mold cavity under controlled conditions. This problem centers on controlling process parameters to minimize part defects, particularly warpage on different walls of the molded component. The goal is to identify optimal injection molding settings that minimize the overall warpage across the four walls of the part. In this example, the low-fidelity data consist of simulation outputs from computational fluid dynamics or process simulation software that predict warpage based on material properties and process conditions. These simulations are fast to execute and can be run at many different parameter settings. The high-fidelity data are experimental measurements of warpage from physical injection molding trials, which accurately reflect the true manufacturing process but are limited in number due to the cost and time required for physical experimentation.

\end{example}

\subsection{Related Literature}

Computer simulations are essential tools for studying complex physical systems, enabling engineers to explore design spaces and optimize processes at relatively low cost compared to physical experimentation \citep{forrester2007multi}. However, a fundamental challenge in using computer models is that they inevitably include simplifications, assumptions, and abstractions. Thus, even with optimal parameter settings, a systematic difference typically remains between model predictions and actual physical behavior, known as model discrepancy. Recognizing and quantifying this discrepancy is critical for obtaining realistic uncertainty quantification and reliable predictions \citep{gramacy2020surrogates}. To address this, the Bayesian statistical framework provides a principled approach to computer model calibration \citep{higdon2004combining}. By treating unknown quantities such as calibration parameters and model discrepancy as random variables with prior probability distributions, Bayesian methods integrate prior information with simulation outputs to yield posterior distributions of uncertainty.

Implementing Bayesian calibration typically requires efficient surrogate models. Gaussian process (GP) regression, also known as Kriging, is widely used for this purpose as it treats simulator outputs as realizations of a GP and provides inherent uncertainty quantification. While traditional GP surrogates assume data come from a single fidelity level, this assumption is often inefficient for complex problems where data sources vary in cost and accuracy \citep{forrester2007multi}. Consequently, multi-fidelity approaches have been developed to reduce computational costs while maintaining accuracy. Early methods used linear regression to model relationships between fidelities \citep{craig2001multifidelity,kennedy2001bayesian}, while more flexible extensions capture these relationships within a GP using autoregressive structures \citep{legratiet2013recursive,gramacy2020surrogates}. The most prominent of these is the Kennedy and O'Hagan (KOH) framework \citep{kennedy2001bayesian}. This framework explicitly models the discrepancy between true physical processes (high-fidelity) and computer simulations (low-fidelity), which allows us to estimate parameters and correct model errors at the same time.

Despite the wide use of the KOH framework, it faces several implementation challenges. First, parameter identifiability is a common challenge, as it is often difficult to distinguish between calibration parameters and model discrepancy \citep{tuo2015efficient,plumlee2017orthogonal}. Second, for complex computer models, computational costs remain a bottleneck. Recent advances have addressed this by introducing enhanced GP methods, such as deep or elastic GPs, to improve flexibility and scalability \citep{marmin2022deep,francom2023elastic}.

Prior research has primarily focused on improving the accuracy of model calibration and prediction. However, the influence of parameter uncertainty on the optimization process is often neglected. In traditional methods like Response Surface Methodology (RSM) \citep{myers2016response}, a single optimal setting is typically derived from a fitted model. This deterministic approach fails to account for the inherent uncertainty within the model parameters. To address this limitation, the concept of decision uncertainty was recently introduced to manufacturing optimization, e.g., \cite{li2026uncertainty}. In contrast to traditional methods that provide only point estimates, this approach utilizes Bayesian inference to generate a distribution of optimal settings. This is essential for understanding the reliability of the recommended parameters considering model uncertainty.

\subsection{Our Contribution}
This paper develops a decision analysis framework built on multi-fidelity GP models that integrate high- and low-fidelity data sources. Our contributions include formulating a surrogate that combines low-fidelity simulation data with high-fidelity experimental observations via a GP structure, extending the KOH framework to include decision analysis capabilities, and integrating the surrogate into a five-stage decision workflow demonstrated through real-world applications.

The remainder of this paper is organized as follows. Section \ref{sec:cali_model} outlines the proposed five-stage workflow for solving generic calibration-based decision problems, detailing the optimization problem and multi-fidelity data structure (Figure \ref{fig:workflow}). Section \ref{subsec:koh_framework} develops the Bayesian calibration framework, covering the joint GP modeling approach and the parameter estimation method. Section \ref{sec:decision_analysis} presents the decision analysis algorithm for identifying optimal input settings under uncertainty. Section \ref{sec:illustrative_example} validates the framework using a synthetic dataset. Finally, Sections \ref{sec:real_study_1} and \ref{sec:real_study_2} demonstrate the complete workflow through two real-world engineering applications.

\section{Problem Overview}
\label{sec:cali_model}

%{\color{red} 1. introduce the optimization problem 2. introduce the model, discuss the data, and the joint normal distribution, 2, estimation and prediction}
%{\color{red}Change $Y^F$ to be $Y^H$ and $Y^M$ to be $Y^L$, representing High Fidelity and Low Fidelity respectively. Can consider follow some set up in the paper of:
%\url{https://josselin-garnier.org/wp-content/uploads/2014/09/recursivecokriging.pdf}}

We consider an optimization problem of the form
\begin{equation}\label{eq:obj}
\min_{\mathbf{x} \in \mathcal{X}} \, G(\mathbf{y}(\mathbf{x})),
\end{equation}
where $\mathbf{y}(\mathbf{x})=(y_1(\mathbf{x}),\ldots, y_p(\mathbf{x}))^\top\in\mathbb{R}^p$ is a vector of outputs, \(\mathbf{x} = (x_1, \dots, x_d)^\top\) denotes a \(d\)-dimensional vector of inputs in the feasible domain \(\mathcal{X} \subset \mathbb{R}^d\), and \(G: \mathbb{R}^p \to \mathbb{R}\) is an objective function defined on the outputs \(\mathbf{y}(\mathbf{x})\). The specific formulation of the objective function $G(\cdot)$ varies across different problems.

We define the system outputs and objective functions for the two examples as follows. In Example \ref{ex1}, the system output is a scalar deformation $y(\mathbf{x})$, and the objective function is simply the deformation itself:
\[
G(y(\mathbf{x}))=y(\mathbf{x}).
\]
In Example \ref{ex2}, the system generates a four dimensional output vector $\mathbf{y}(\mathbf{x})=(y_1(\mathbf{x}), \dots, y_4(\mathbf{x}))^\top$, representing the displacement of four walls. The objective function aims to reduce the overall displacement:
\[
G(\mathbf{y}(\mathbf{x}))=\sum^4_{k=1}y^2_k(\mathbf{x}).
\]

For both examples, obtaining high-fidelity data is computationally or experimentally expensive. To reduce decision uncertainty under budget constraints, we integrate low-fidelity data from alternative sources. Specifically, in Example \ref{ex1}, the low-fidelity data are derived from efficient DoC models, while the high-fidelity data come from computationally expensive full stress-deformation simulations. In Example \ref{ex2}, the low-fidelity data consist of warpage simulation predictions, whereas the high-fidelity data are obtained from physical injection molding trials.

\begin{figure}[H]
  \centering
  \includegraphics[width=1\textwidth]{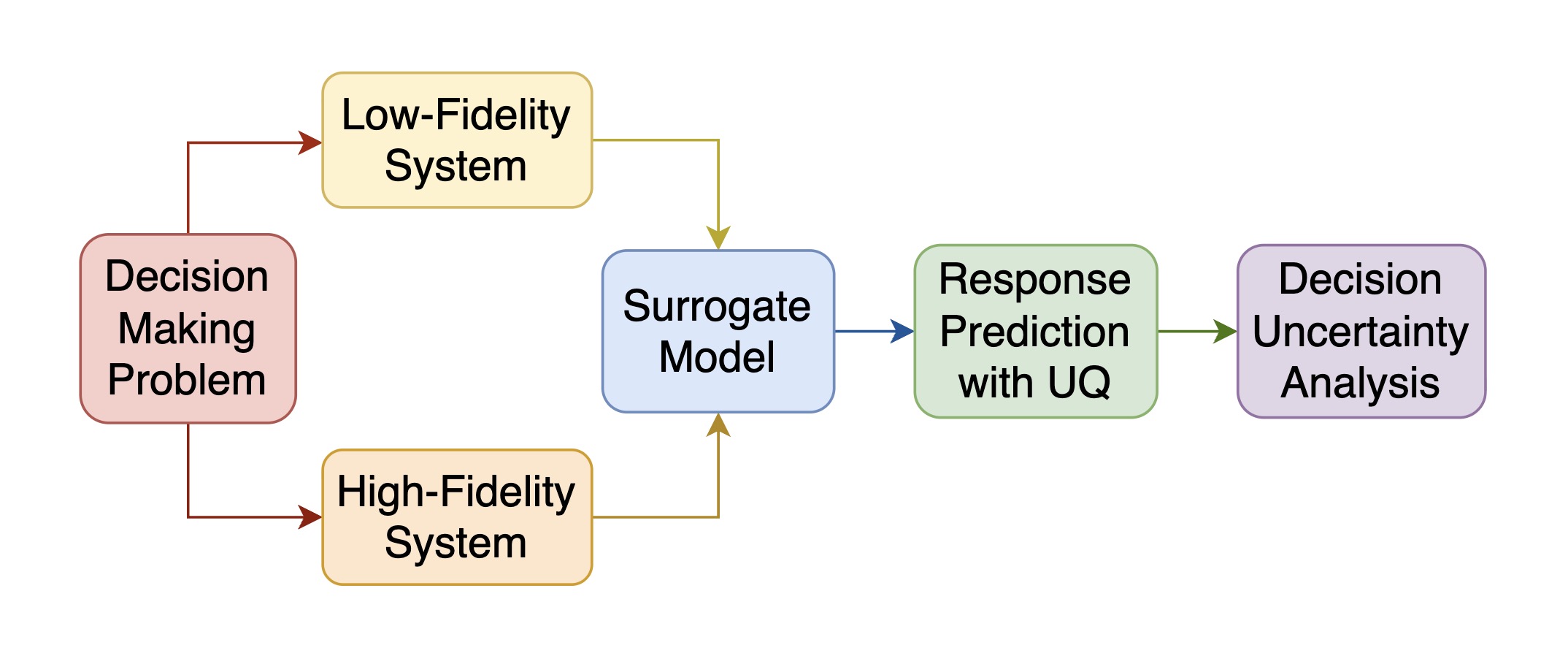}
  \caption{Decision Analysis Framework for High-fidelity and Low-fidelity Systems}
  \label{fig:workflow}
\end{figure}

Let $\mathbf{y}^H(\mathbf{x})$ and $\mathbf{y}^L(\mathbf{x})$ represent the outcome vectors corresponding to the high-fidelity and low-fidelity data sources, respectively. Both serve as approximations of the true system outcome vector $\mathbf{y}(\mathbf{x})$. Our goal is to leverage both sources of information, $\mathbf{y}^L(\mathbf{x})$ and $\mathbf{y}^H(\mathbf{x})$, to efficiently approximate $\mathbf{y}(\mathbf{x})$ and then search for the minimizer of the objective $G$ in \eqref{eq:obj}. We employ a statistical surrogate model to combine the inexpensive low-fidelity outputs $\mathbf{y}^L(\mathbf{x})$ with the expensive high-fidelity outputs $\mathbf{y}^H(\mathbf{x})$ to improve the prediction accuracy of \(\mathbf{y}(\mathbf{x})\) across \(\mathcal{X}\) and subsequently perform decision analysis with respect to the objective function in \eqref{eq:obj}. The overall framework is illustrated in Figure \ref{fig:workflow}.

\section{Multi-fidelity Bayesian Calibration: Modeling, Estimation, and Prediction}
\label{subsec:koh_framework}

\subsection{Joint Gaussian Process Model}
\label{subsec:koh_framework_concise}

For notational convenience, we describe our model under the special case that the dimension of the output $p=1$, i.e., the output $y(\mathbf x)$ is a scalar. We will illustrate how our model can be used for a vector outcome in a case study.  We model the dependency between high-fidelity and low-fidelity outputs through a calibration structure that explicitly accounts for potential discrepancy. This approach follows the classic KOH calibration model \citep{kennedy2001bayesian} and multi-fidelity Gaussian process extensions, such as recursive co-kriging \citep{legratiet2013recursive}.
%({\color{red} also refer to RECURSIVE CO-KRIGING MODEL}).
We assume that the high-fidelity model $y^H(\mathbf{x})$ can be expressed as a scaled version of the low-fidelity model plus a systematic discrepancy term:
\begin{equation}
y^H(\mathbf{x}) = u \cdot y^L(\mathbf{x}) + b(\mathbf{x})
\label{eq:multifidelity_relationship}
\end{equation}
where $u$ is the unknown calibration parameter that scales the low-fidelity output, and $b(\mathbf{x})$ represents the systematic discrepancy between the scaled low-fidelity model and the high-fidelity model. 
For noisy high-fidelity observations, we model:
\begin{equation}
Y^H(\mathbf{x}) = y^H(\mathbf{x}) + \varepsilon = u \cdot y^L(\mathbf{x}) + b(\mathbf{x}) + \varepsilon
\label{eq:observation_model}
\end{equation}
where $\varepsilon \sim \mathcal{N}(0, \sigma_\varepsilon^2)$ represents independent Gaussian measurement errors across different experimental runs.

Following the KOH model, we surrogate both $y^L(\mathbf{x})$ and $b(\mathbf{x})$ using the GP model. Assume that
\begin{align}
y^L(\mathbf{x}) &\sim \mathcal{GP}(\mu_L(\mathbf{x}), \Sigma_L(\mathbf{x}, \mathbf{x}')) \label{eq:gp_prior_low} \\
b(\mathbf{x}) &\sim \mathcal{GP}(0, \Sigma_b(\mathbf{x}, \mathbf{x}')) \label{eq:gp_prior_discrepancy}
\end{align}
For the low-fidelity model, we use a mean function $\mu_L(\mathbf{x})$ to incorporate prior knowledge about the system behavior, and a covariance function $\Sigma_L(\mathbf{x}, \mathbf{x}')$ that captures the correlation structure across inputs. For the discrepancy function, we assume a zero mean GP prior, reflecting our assumption that the scaled low-fidelity model captures the global trend of the high-fidelity process, with remaining deviations modeled by the covariance structure.
We employ the separable power exponential covariance functions \citep{santner2003design} for both GPs:
\begin{align}
\Sigma_L(\mathbf{x}, \mathbf{x}') &= \sigma_L^2 \exp\left(-\frac{1}{2}\sum_{i=1}^d \frac{(x_i - x_i')^2}{\ell_{L,i}^2}\right) \label{eq:cov_low} \\
\Sigma_b(\mathbf{x}, \mathbf{x}') &= \sigma_b^2 \exp\left(-\frac{1}{2}\sum_{i=1}^d \frac{(x_i - x_i')^2}{\ell_{b,i}^2}\right) \label{eq:cov_discrepancy}
\end{align}
where $\sigma_L^2, \sigma_b^2$ are variance parameters, and $\ell_{L,i}, \ell_{b,i}$ are length-scale parameters for each input dimension.

Given the hierarchical structure defined in \eqref{eq:multifidelity_relationship} and \eqref{eq:observation_model}, along with the GP priors specified in \eqref{eq:gp_prior_low} and \eqref{eq:gp_prior_discrepancy}, the high-fidelity and low-fidelity outputs $(y^L(\mathbf x),y^H(\mathbf x))^\top$ follow a joint GP \citep{kennedy2001bayesian} with the covariance function defined as:
\[
\mathrm{cov}\left( \begin{pmatrix} y^L(\mathbf{x}) \\ y^H(\mathbf{x}) \end{pmatrix}, \begin{pmatrix} y^L(\mathbf{x}') \\ y^H(\mathbf{x}') \end{pmatrix} \right) =
\begin{pmatrix}
\Sigma_L(\mathbf{x}, \mathbf{x}') & u \Sigma_L(\mathbf{x}, \mathbf{x}') \\
u \Sigma_L(\mathbf{x}, \mathbf{x}') & u^2 \Sigma_L(\mathbf{x}, \mathbf{x}') + \Sigma_b(\mathbf{x}, \mathbf{x}')
\end{pmatrix}.
\]
At a single input location $\mathbf{x}$, their joint distribution follows a bivariate Gaussian distribution: 
\begin{equation}
\begin{pmatrix}
y^L(\mathbf{x}) \\
y^H(\mathbf{x})
\end{pmatrix} \sim \mathcal{N}\left(
\begin{pmatrix}
\mu_L(\mathbf{x}) \\
u \mu_L(\mathbf{x})
\end{pmatrix},
\begin{pmatrix}
\Sigma_L(\mathbf{x}, \mathbf{x}) & u \Sigma_L(\mathbf{x}, \mathbf{x}) \\
u \Sigma_L(\mathbf{x}, \mathbf{x}) & u^2 \Sigma_L(\mathbf{x}, \mathbf{x}) + \Sigma_b(\mathbf{x}, \mathbf{x})
\end{pmatrix}
\right)
\label{eq:joint_distribution}
\end{equation}

Given input and output data for both systems, the stacked outputs follow a multivariate normal distribution based on the model assumption in \eqref{eq:multifidelity_relationship}.
Let \(\mathbf{X}_L = \{\mathbf{x}_1^L, \ldots, \mathbf{x}_{n_L}^L\}\) denote the set of \(n_L\) low-fidelity inputs, where each $\mathbf{x}_i^L$ is a vector in $\mathbb{R}^d$. Let
\(\mathbf{Y}^L = (Y_1^L, \ldots, Y_{n_L}^L)^\top
\), where \(Y_i^L=y^L(\mathbf{x}_i^L)\), be the corresponding low-fidelity outputs. Similarly, let \(\mathbf{X}_H = \{\mathbf{x}_1^H, \ldots, \mathbf{x}_{n_H}^H\}\) denote the set of \(n_H\) high-fidelity input locations, and let
$\mathbf{Y}^H = (Y_1^H, \ldots, Y_{n_H}^H)^\top$ be the outputs of the high-fidelity model with noisy observations
\[
Y_j^H = y^H(\mathbf{x}_j^H) + \varepsilon_j,
\quad
\varepsilon_j \sim \mathcal{N}(0, \sigma_\varepsilon^2).
\]
Typically, \(n_L \gg n_H\) because low-fidelity evaluations are much cheaper to obtain. The high-fidelity design points \(\mathbf{X}_H\) may be a subset of \(\mathbf{X}_L\), although this is not required.

Let $\mathbf{Y}=((\mathbf{Y}^L)^\top, (\mathbf{Y}^H)^\top)^\top$
be the output vector that stacks $\mathbf{Y}^L$ and $\mathbf{Y}^H$ together. 
Under our model assumptions, the joint distribution of all observations is multivariate Gaussian:
\[
\mathbf{Y} \sim \mathcal{N}(\boldsymbol{\mu}, \mathbf{\Sigma})
\]
where the mean vector is:
\[
\boldsymbol{\mu} = \begin{pmatrix} \boldsymbol{\mu}_L(\mathbf{X}_L) \\ u \cdot \boldsymbol{\mu}_L(\mathbf{X}_H) \end{pmatrix}
\]
and the covariance matrix has the block structure:
\[
\mathbf{\Sigma} = \begin{pmatrix}
\mathbf{\Sigma}_{LL} & u \cdot \mathbf{\Sigma}_{LH} \\
u \cdot \mathbf{\Sigma}_{LH}^T & u^2 \cdot \mathbf{\Sigma}_{HH} + \mathbf{\Sigma}_{bb} + \sigma_\varepsilon^2 \mathbf{I}_{n_H}
\end{pmatrix}
\]

Here, the sub-matrices are constructed element-wise from the covariance functions. Specifically, $\mathbf{\Sigma}_{LL}$ denotes the covariance matrix of the low-fidelity inputs, where the $(i,j)$-th entry is $\Sigma_L(\mathbf{x}_i^L, \mathbf{x}_j^L)$. The matrix $\mathbf{\Sigma}_{LH}$ represents the cross-covariance between low- and high-fidelity inputs, with entries $\Sigma_L(\mathbf{x}_i^L, \mathbf{x}_j^H)$. It is important to note that $\mathbf{\Sigma}_{HH}$ is also computed using the low-fidelity covariance function $\Sigma_L(\cdot, \cdot)$, but evaluated at the high-fidelity input locations $\mathbf{X}_H$. The matrix $\mathbf{\Sigma}_{bb}$ represents the covariance of the discrepancy function evaluated at $\mathbf{X}_H$, with entries $\Sigma_b(\mathbf{x}_i^H, \mathbf{x}_j^H)$. The term $\sigma_\varepsilon^2 \mathbf{I}_{n_H}$ accounts for independent observation noise in the high-fidelity data. 
\subsection{Parameter Estimation}
\label{subsec:param_estimation}

We provide parameter estimation procedure following \citep{gramacy2020surrogates}.
First, we fit a GP emulator to the low-fidelity data \(\{\mathbf{X}_L,\mathbf{Y}^L\}\). The hyperparameters \(\hat{\boldsymbol\theta}_L = \{\hat{\sigma}_L^2, \hat{\ell}_{L,1},\ldots,\hat{\ell}_{L,d}\}\) are estimated via maximum likelihood estimation based on the covariance structure \(\Sigma_L\) defined in Section \ref{subsec:koh_framework_concise}. This fitted GP emulator provides the predictions \(\hat{y}^L(\mathbf{x})\) that will be used to predict the low-fidelity response at the high-fidelity input locations.

Next, for a given candidate value of the calibration parameter \(u\), we treat the difference between the high-fidelity observations and the scaled low-fidelity predictions as the realization of the discrepancy process plus noise. Based on the discrepancy prior $b(\mathbf{x}) \sim \mathcal{GP}(0, \Sigma_b)$, the marginal likelihood function for the data is derived from the multivariate normal distribution:
\[
\mathbf{Y}^H - u \cdot \hat{y}^L(\mathbf{X}_H) \sim \mathcal{N}(\mathbf{0}, \mathbf{\Sigma}_{bb} + \sigma_\varepsilon^2 \mathbf{I}),
\]
where $\mathbf{\Sigma}_{bb}$ is the covariance matrix of the discrepancy process at high-fidelity locations. We then estimate the discrepancy hyperparameters \(\hat{\boldsymbol\theta}_{b|u}\) by maximizing this likelihood.

The optimal calibration parameter \(\hat{u}\) is found by maximizing the logarithmic marginal likelihood of the discrepancy GP combined with a prior on \(u\):
\[
\hat{u} = \arg\max_u \left\{ \log p(\mathbf{Y}^H - u \cdot \hat{y}^L(\mathbf{X}_H) \mid \mathbf{X}_H, \hat{\boldsymbol\theta}_{b|u}) + \log p(u) \right\}.
\]
where \(p(u)\) is the prior distribution of the calibration parameter.

 While the optimization above provides a point estimate, it does not explicitly quantify the parameter uncertainty. To address this, we employ a Leave-One-Out (LOO) cross-validation procedure. Specifically, we iteratively hold out each high-fidelity observation and re-estimate \(u\) using the remaining \(n_H-1\) data points. The resulting empirical distribution of these \(n_H\) estimates serves as a data driven approximation of the posterior distribution of \(u\), capturing its variability given the limited high-fidelity data.

\subsection{Posterior Predictive Distribution}
\label{subsec:posterior_prediction}

Given the estimated calibration parameter $\hat{u}$ and the fitted hyperparameters, we derive the posterior predictive distribution for the high-fidelity response at any new input location $\mathbf{x}_{new} \in \mathcal{X}$.

Based on the joint Gaussian structure constructed in Section \ref{subsec:koh_framework_concise}, the conditional distribution of the high-fidelity process given the observed outcome $\mathbf{Y}$ follows a GP:
\begin{equation}
y^H(\mathbf{x}_{new}) \mid \mathbf{Y}, \hat{u} \sim \mathcal{GP}\left( \mu_{new}(\mathbf{x}_{new}), \Sigma_{new}(\mathbf{x}_{new}, \mathbf{x}_{new}') \right).
\label{eq:posterior_gp}
\end{equation}
The posterior mean function $\mu_{new}$ and covariance function $\Sigma_{new}$ are derived using standard conditional Gaussian identities:
\begin{align}
\mu_{new}(\mathbf{x}_{new}) &= \hat{u}\mu_L(\mathbf{x}_{new}) + \mathbf{k}(\mathbf{x}_{new})^\top \mathbf{\Sigma}^{-1} (\mathbf{Y} - \boldsymbol{\mu}), \\
\Sigma_{new}(\mathbf{x}_{new}, \mathbf{x}_{new}') &= V(\mathbf{x}_{new}, \mathbf{x}_{new}') - \mathbf{k}(\mathbf{x}_{new})^\top \mathbf{\Sigma}^{-1} \mathbf{k}(\mathbf{x}_{new}').
\end{align}
Here, $\mathbf{\Sigma}$ is the joint covariance matrix of the observations defined in Section \ref{subsec:koh_framework_concise}. The term $\mathbf{k}(\mathbf{x}_{new})$ represents the covariance vector between the observed data $\mathbf{Y}$ and the prediction at $\mathbf{x}_{new}$, constructed as:
\[
\mathbf{k}(\mathbf{x}_{new}) = \begin{pmatrix} \hat{u} \cdot \Sigma_L(\mathbf{X}_L, \mathbf{x}_{new}) \\ \hat{u}^2 \cdot \Sigma_L(\mathbf{X}_H, \mathbf{x}_{new}) + \Sigma_b(\mathbf{X}_H, \mathbf{x}_{new}) \end{pmatrix}.
\]
The term $V(\mathbf{x}_{new}, \mathbf{x}_{new}')$ represents the prior variance structure at the new location:
\[
V(\mathbf{x}_{new}, \mathbf{x}_{new}') = \hat{u}^2 \cdot \Sigma_L(\mathbf{x}_{new}, \mathbf{x}_{new}') + \Sigma_b(\mathbf{x}_{new}, \mathbf{x}_{new}').
\]

\section{Decision Analysis}
\label{sec:decision_analysis}

We now return to the original problem formulation in \eqref{eq:obj}, where the system output is a vector $\mathbf{y}^H(\mathbf{x}) \in \mathbb{R}^p$. Since the true functions $y_k^H(\mathbf{x})$ ($k=1,\dots,p$) are unknown, we model them as Gaussian Processes. For multi-output problems, we apply the posterior prediction derived in Equation \eqref{eq:posterior_gp} to each output dimension independently. Consequently, the conditional distribution for the $k$-th dimension on a set of candidate inputs $\mathbf{X}_{cand} \in \mathbb{R}^{N \times d}$ follows a Multivariate Normal distribution:
\begin{equation}
\mathbf{y}_k^H(\mathbf{X}_{cand}) \mid \mathbf{Y}_k, \hat{u}_k \sim \mathcal{N}(\boldsymbol{\mu}_{new,k}(\mathbf{X}_{cand}), \mathbf{\Sigma}_{new,k}(\mathbf{X}_{cand})),
\label{eq:predictive_distribution}
\end{equation}
where $\boldsymbol{\mu}_{new,k}$ and $\mathbf{\Sigma}_{new,k}$ are the posterior mean vector and covariance matrix for the $k$-th output.

To quantify the decision uncertainty, we employ a simulation-based approach. We draw independent samples from these distributions for each dimension, calculate the objective function value, and identify the candidate point that minimizes it. This procedure is detailed in Algorithm \ref{alg:decision_analysis}.

\begin{algorithm}[H]
\caption{Bayesian Decision Analysis for Optimal Input Identification}
\label{alg:decision_analysis}
\begin{algorithmic}
\REQUIRE Posterior samples of calibration parameters $\{\mathbf{u}^{(s)}\}_{s=1}^{N_u}$ (where $N_u$ is the sample size and $\mathbf{u}^{(s)}$ contains parameters for all $p$ outputs), low-fidelity emulators, high-fidelity data, replications $n_{rep}$, number of candidates $N$
\ENSURE Optimal input locations $\mathcal{X}_{opt}$
\STATE $\mathcal{X}_{opt}\leftarrow\emptyset$
\FOR{$s=1$ \TO $N_u$}
    \STATE Update discrepancy GP hyperparameters for all $p$ outputs conditional on $\mathbf{u}^{(s)}$
    \FOR{$r=1$ \TO $n_{rep}$}
        \STATE Generate candidate set $\mathbf{X}_{cand} \in \mathbb{R}^{N \times d}$ via Latin Hypercube Design
        \FOR{$k=1$ \TO $p$}
            \STATE Compute $(\boldsymbol{\mu}_{new,k}, \mathbf{\Sigma}_{new,k})$ on $\mathbf{X}_{cand}$ for output $k$
            \STATE Draw realization $\mathbf{y}_{samp,k} \sim \mathcal{N}(\boldsymbol{\mu}_{new,k}, \mathbf{\Sigma}_{new,k})$
        \ENDFOR
        \STATE $\mathbf{x}_{s,r}^{*} \leftarrow \underset{\mathbf{x}_m \in \mathbf{X}_{cand}}{\arg\min} G\left( \mathbf{y}_{samp,1}(\mathbf{x}_m), \dots, \mathbf{y}_{samp,p}(\mathbf{x}_m) \right)$
        \STATE $\mathcal{X}_{opt}\leftarrow\mathcal{X}_{opt}\cup\{\mathbf{x}_{s,r}^{*}\}$
    \ENDFOR
\ENDFOR
\RETURN $\mathcal{X}_{opt}$
\end{algorithmic}
\end{algorithm}

In each iteration, we generate $N$ candidate input points arranged in a matrix $\mathbf{X}_{cand}$, compute the posterior parameters, and draw a realization $\mathbf{y}_{samp}$ from the multivariate normal distribution. We then identify the specific row in $\mathbf{X}_{cand}$ that minimizes the objective function of this realization.

The resulting collection of optimal inputs across all $N_u$ posterior samples and $n_{rep}$ replications defined as
\begin{equation}
\mathcal{X}_{opt} = \{\mathbf{x}_{s,r}^{*} : s=1,\ldots,N_u, \, r=1,\ldots,n_{rep}\}
\label{eq:xopt_collection}
\end{equation}
provides an empirical approximation of the posterior distribution of the optimal inputs. We can then analyze this distribution by plotting histograms for each variable of $\mathbf{x}$. The medians of these distributions suggest the most probable input settings that minimize the squared response. The spread of these distributions quantifies the decision uncertainty. Wide or multi-modal distributions indicate that the optimum is not well-defined or that multiple input regions may lead to similar optimal outcomes.

\section{Illustrative Example}
\label{sec:illustrative_example}

We now demonstrate the proposed decision analysis framework on an illustrative example with two input variables, $\mathbf{x} = (x_1, x_2) \in [-1, 1]^2$. 
In this example, our objective is simply to minimize the response of the high-fidelity process. Thus, the specific problem instance is:
\begin{equation}
\min_{\mathbf{x} \in [-1, 1]^2} \, y^H(\mathbf{x}).
\end{equation}

\subsection{Data Generation}
\label{subsec:computer_model_data}
To evaluate the method's ability to handle structural bias, we generate data from two distinct functions. 
The low-fidelity model $y^L(\mathbf{x})$ is defined as:
\begin{equation}
y^L(\mathbf{x}) = (x_1 - a_{L1})^2 + (x_2 - a_{L2})^2,
\label{eq:computer_model_true}
\end{equation}
where the minimum is located at $a_L = (a_{L1}, a_{L2}) = (-0.6, 0.2)$. We generate $n_L = 200$ samples using a Latin Hypercube Design (LHD).

The true high-fidelity process $y^H(\mathbf{x})$ is defined as:
\begin{equation}
y^H(\mathbf{x}) = (x_1 - a_{H1})^2 + (x_2 - a_{H2})^2,
\label{eq:physical_process_true}
\end{equation}
where the true minimum is at $a_H = (a_{H1}, a_{H2}) = (-0.8, 0.4)$. 
We generate sparse noisy observations $\mathbf{Y}^H$ at $n_H = 50$ points:
\begin{equation}
\mathbf{Y}^H = y^H(\mathbf{x}) + \varepsilon, \quad \varepsilon \sim \mathcal{N}(0, \sigma_\varepsilon^2) \text{ with } \sigma_\varepsilon = 0.02.
\label{eq:field_observation_generated}
\end{equation}

It is crucial to note that we intentionally set $a_L \neq a_H$. This discrepancy simulates a common manufacturing scenario where the computer model ($y^L$) is precise (low variance) but structurally biased. Relying solely on $y^L$ would lead to a confident but incorrect decision at $a_L$, while relying solely on sparse $Y^H$ data would lead to high uncertainty. Our goal is to recover the true minimum $a_H$ by combining both sources.

\subsection{Parameter Estimation Results}
The multi-fidelity GP model described in Section \ref{subsec:koh_framework_concise} was applied to the generated datasets $\mathbf{Y}^L$ and $\mathbf{Y}^H$. Since the functional forms are similar, we expect the scaling parameter $u \approx 1$, with the discrepancy term $b(\mathbf{x})$ capturing the shift in minima location.
Using the LOO procedure mentioned in Section \ref{subsec:param_estimation}, we estimated the calibration parameter $\hat{u}= 1.0958$ with a narrow 95\% confidence interval of $(1.0844, 1.0958)$. This indicates a strong correlation between the fidelities despite the structural bias.

\subsection{Decision Analysis}
We generated $N_u = 100$ posterior samples of $u$ and, for each, generated $n_{rep}=100$ predictive surfaces to find the optimal inputs minimizing $y^H$. This process yielded a collection of $10,000$ potential optima, corresponding to the set $\mathcal{X}_{opt}$ defined in Equation \eqref{eq:xopt_collection}.
We compare three scenarios to demonstrate the necessity of the multi-fidelity approach.

\begin{figure}[H]
  \centering
  \includegraphics[width=0.8\textwidth]{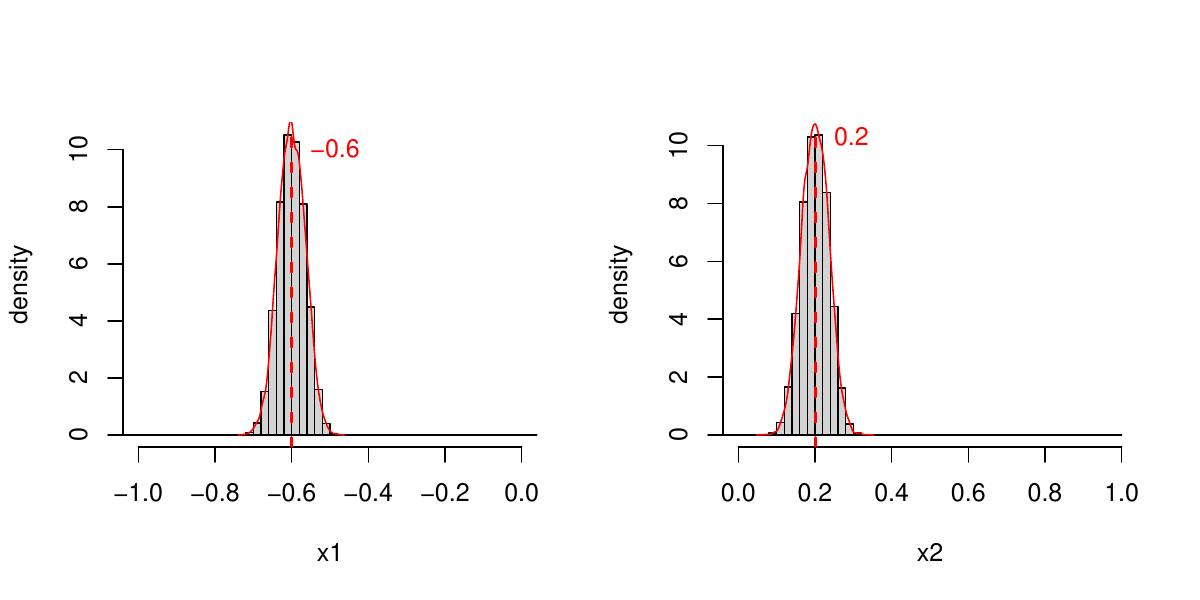}
  \caption{Distribution of minima locations obtained by optimizing the GP emulator fitted only to the low-fidelity data.}
  \label{fig:simu_MM}
\end{figure}

\textbf{Scenario 1: Low-Fidelity Only.}
Figure \ref{fig:simu_MM} shows the distribution of minima using only $\mathbf{Y}^L$. The results are highly concentrated (standard deviation $\approx 0.035$), indicating high precision. However, the median location is $(-0.6, 0.2)$, which aligns with the biased model minimum $a_L$ but is far from the true physical minimum $a_H$. This confirms that low-fidelity data alone leads to precise but incorrect decisions.

\begin{figure}[H]
  \centering
  \includegraphics[width=0.8\textwidth]{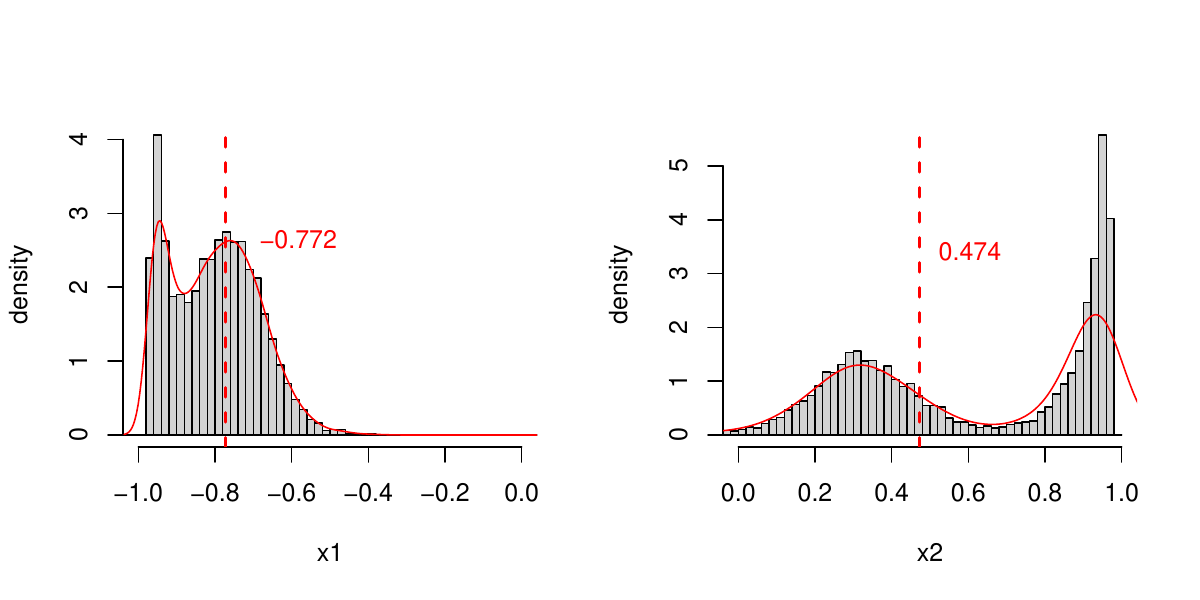}
  \caption{Distribution of minima locations obtained by optimizing the GP model fitted only to the sparse high-fidelity observations.}
  \label{fig:simu_FF}
\end{figure}

\textbf{Scenario 2: High-Fidelity Only.}
Figure \ref{fig:simu_FF} shows the results using only sparse high-fidelity observations. The median location $(-0.772, 0.474)$ is close to the truth $a_H$, indicating that the method captures the correct region. However, due to the sparse and noisy nature of the data, the standard deviations are unacceptably large ($0.628$ for $x_1$ and $0.422$ for $x_2$). The wide spread indicates that the optimal solution cannot be reliably identified.

\begin{figure}[H]
  \centering
  \includegraphics[width=0.8\textwidth]{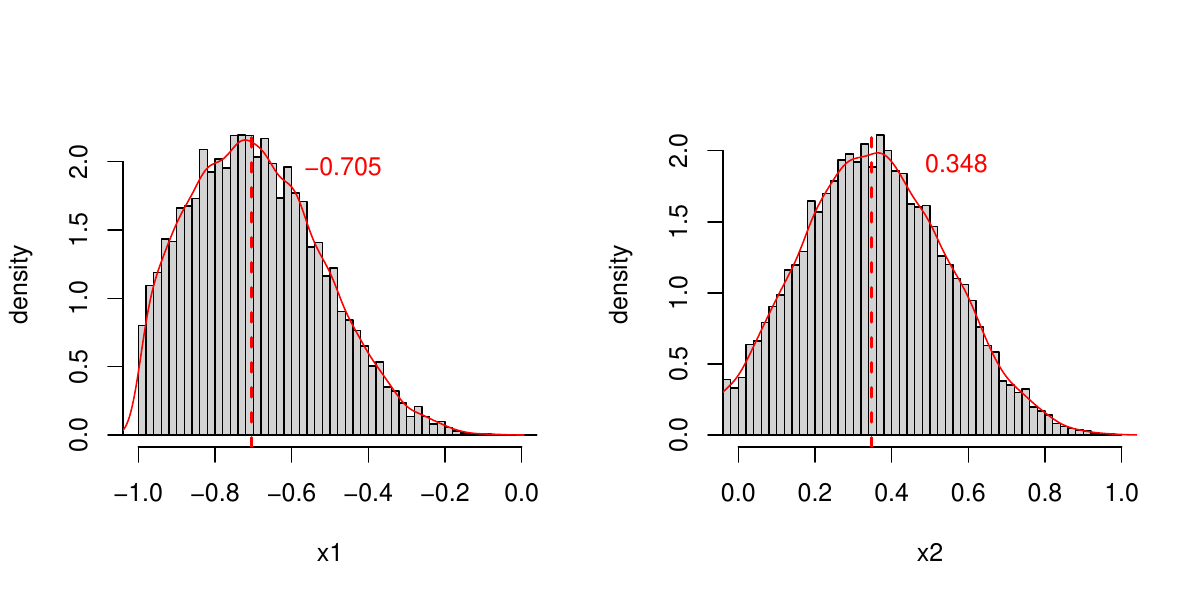}
  \caption{Distribution of minima locations obtained by applying Bayesian calibration to both low-fidelity data and high-fidelity observations.}
  \label{fig:simu_MF}
\end{figure}

\textbf{Scenario 3: Multi-Fidelity Calibration (The Solution).}
Figure \ref{fig:simu_MF} presents the distribution of the optimal candidates in $\mathcal{X}_{opt}$ obtained by our proposed framework. The median location shifts to $(-0.705, 0.348)$, effectively correcting the bias from the low-fidelity model. Simultaneously, the standard deviations decrease to $0.170$ and $0.193$, significantly reducing uncertainty compared to Scenario 2. This demonstrates that our framework effectively combines the precision of simulations with the accuracy of physical experiments.

\subsection{Robustness}

To further evaluate the robustness and stability of our proposed framework, we repeated the entire data generation and optimization process on 50 independent datasets. For each dataset $j$ ($j=1,\dots,50$), we identified the robust optimal solution by taking the median of the optimal candidates in $\mathcal{X}_{opt}$ (Equation \ref{eq:xopt_collection}), denoted as $\tilde{x}_{l}^{*(j)}$. We then assessed the performance using the Mean Squared Error (MSE) across these 50 datasets:

\begin{equation}
\text{MSE}_l = \frac{1}{50} \sum_{j=1}^{50} \left( \tilde{x}_{l}^{*(j)} - a_{Hl} \right)^2, \quad l=1,2
\label{eq:mse_formula}
\end{equation}
where $\tilde{x}_{l}^{*(j)}$ is the median of the posterior optimal inputs for the $l$-th variable in the $j$-th dataset.

\begin{figure}[H]
  \centering
  \includegraphics[width=0.8\textwidth]{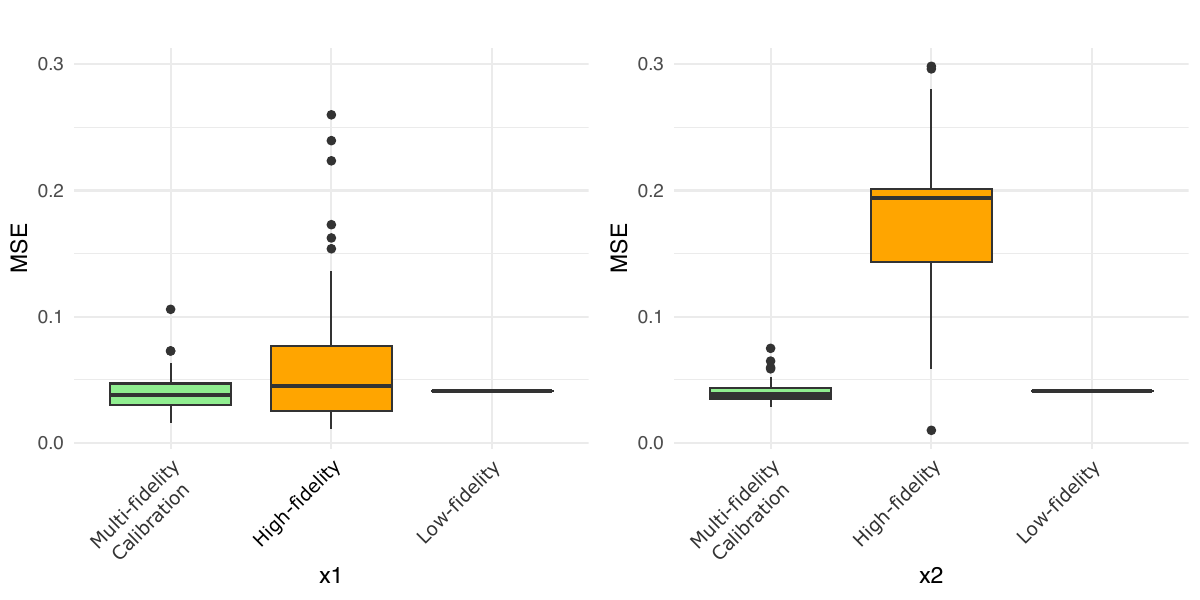}
  \caption{The distributions of MSE values across the 50 datasets.}
  \label{fig:mse_comparison}
\end{figure}

Figure \ref{fig:mse_comparison} presents the MSE distributions across the 50 datasets. The low-fidelity approach exhibits stable but non-zero MSE due to its inherent bias. The high-fidelity approach shows extreme instability with high MSE variance. In contrast, our multi-fidelity approach maintains low and stable MSE values, quantitatively confirming that the proposed method consistently identifies the optimal region with high confidence.

\section{Case Study I: Curing Process Simulation}
\label{sec:real_study_1}

In this section, we consider the curing process simulation described in \citep{limaye2025numericalsimulationinformedrapid} with a commercial finite element analysis software ABAQUS \citep{abaqus2024} and the cure simulation tool COMPRO \citep{compro2024}. The curing process represents a critical stage in the manufacturing of structural components, during which controlled temperature cycles serve as key process parameters. The goal is to minimize deformation output $y$ with respect to four-dimensional inputs representing two critical change points on the temperature-time cycle. This simulation contains two stages: the first stage thermo-chemical analysis generates Degree of Cure (DoC), and the second stage combines DoC with the original four-dimensional inputs for stress-deformation and mechanical analysis to produce the final deformation output $y$. The second stage simulation is more complex and computationally expensive than the first stage. Since second stage outputs depend heavily on first stage outputs, we treat the first stage (DoC) simulation as the low-fidelity system and the complete two-stage simulation as the high-fidelity system.

To facilitate the extensive sampling required for decision analysis, we constructed surrogate environments based on the original simulation data. We fitted second-order polynomial regression models with stepwise variable selection to the original data. Using these models as generators, we created a high-fidelity dataset of $n_H=50$ points and a low-fidelity dataset of $n_L=200$ points via LHD. Gaussian noise with a standard deviation of 0.1 was added to the high-fidelity predictions to simulate measurement variability, while low-fidelity predictions were kept noise-free. The four-dimensional input parameters represent: initial temperature $T_1$, final temperature $T_2$, ramp time $t_1$, and hold time $t_2$. The low-fidelity output corresponds to the dimensionless DoC, while the high-fidelity output measures the deformation magnitude.

We apply the proposed multi-fidelity Bayesian calibration framework to identify optimal cure settings while quantifying decision uncertainty. Following the modular calibration procedure described in Section \ref{subsec:param_estimation}, we obtained a point estimate of $\hat{u} = 9.9535$ with a 95\% confidence interval of $(9.9216, 9.9999)$. This estimated value reflects the physical scaling relationship between the dimensionless DoC (low-fidelity) and the deformation magnitude (high-fidelity). The narrow confidence interval indicates a precise estimation of this relationship and suggests strong consistency between the full simulation and the DoC-only model.

Next, we applied the decision analysis framework described in Algorithm \ref{alg:decision_analysis} to identify optimal input settings that minimize the deformation response. We generated $N_u = 100$ samples from the approximate posterior distribution of $u$. For each sampled calibration parameter $\mathbf{u}^{(s)}$ ($s = 1, \ldots, N_u$), we independently performed $n_{rep} = 100$ optimization replications. In each replication, we randomly generated $N=200$ candidate points using LHD to form the candidate matrix $\mathbf{X}_{cand}$. Using the posterior predictive distributions from Section \ref{subsec:posterior_prediction}, we drawn realizations of the deformation response for each candidate point and identified the input setting that minimized the objective function $G(y) = y$. This procedure, repeated over $N_u \times n_{rep} = 10,000$ iterations, yielded a collection of optimal input locations $\mathcal{X}_{opt}$ that characterizes the decision uncertainty.

\begin{figure}[H]
\centering
\includegraphics[width=0.8\textwidth]{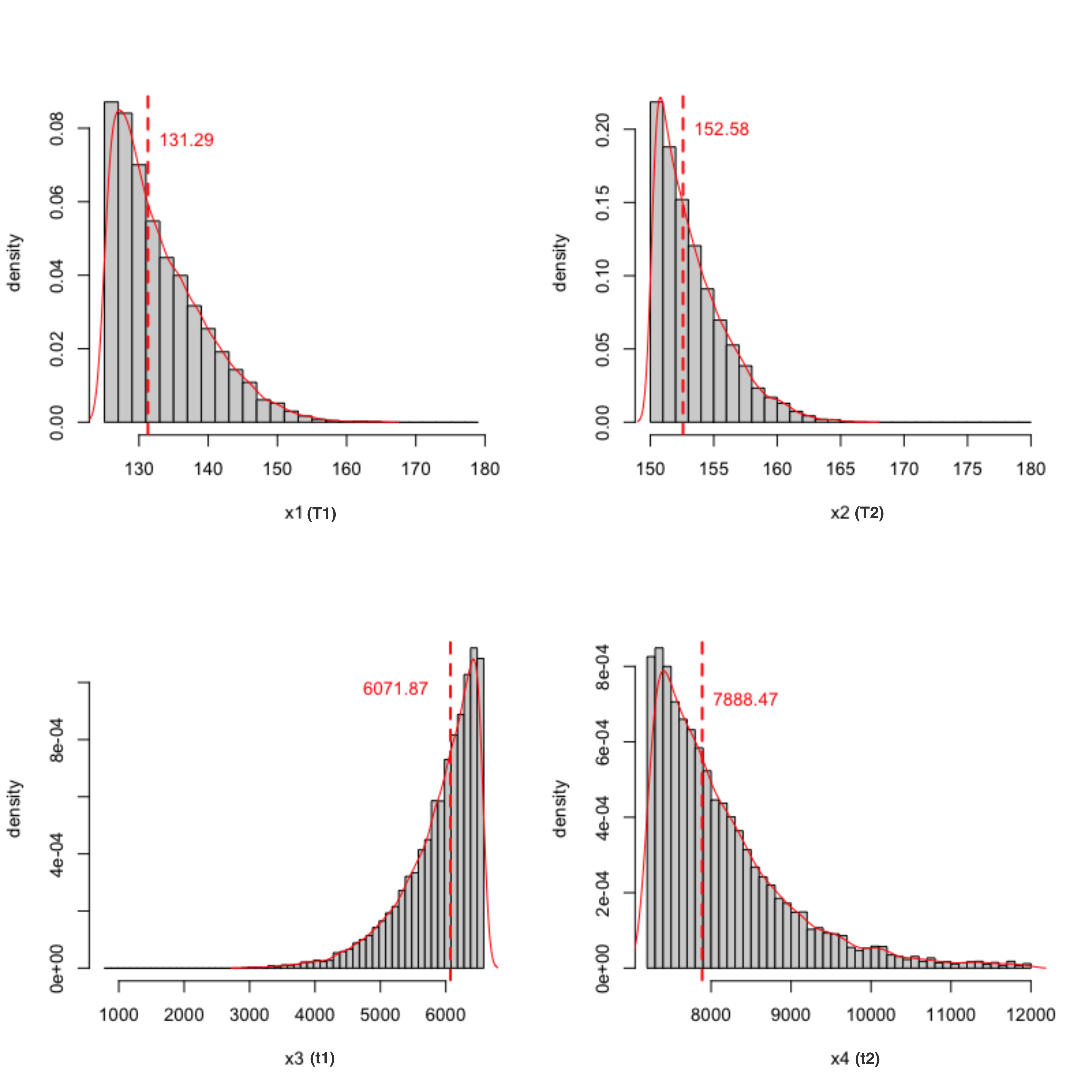}
\caption{Distribution of identified minima locations.}
\label{fig:case1_minima}
\end{figure}

Figure \ref{fig:case1_minima} displays the distributions of the identified optimal locations. The results show that the calibrated optima are tightly concentrated, suggesting that the identified optimal settings are robust to the remaining parameter uncertainty. This consistency demonstrates that, when properly calibrated, the low-fidelity DoC simulation effectively captures the main trends of the deformation behavior to support reliable process optimization.

\section{Case Study II: Injection Molding}
\label{sec:real_study_2}

Injection molding is an important manufacturing process. However, reducing the warpage of injection molding products is still an open challenge. 
In this section, we consider the box-shaped geometry example in \cite{li2026uncertainty}. This system contains four parameters (mold temperature, injection speed, packing pressure and packing time) as the inputs, and generates a four-dimensional output vector $\bm y=(y_1, y_2, y_3, y_4)^\top$ representing the displacement on the four walls of the box-shaped geometry. The objective is to minimize the squared sum of the four-dimensional output. The physical experiments are performed at 27 unique design points, each with three repetitions. 
%{\color{yellow}The real outputs are generated from physical experiments...elaborate with some detail (experimental cost etc) and reference 2-3 sentences}. 
The simulation of injection molding described in \cite{li2026uncertainty} serves as the low-fidelity system to the physical experiments.

In this study, we used 57 low-fidelity simulation runs obtained from injection molding simulations in Moldex3D. For the high-fidelity data, we conducted physical injection molding experiments at 27 carefully selected input design points, with 3 replications per design point, resulting in 81 high-fidelity experimental observations in total. Following the procedure in Section \ref{subsec:koh_framework_concise}, 
we fit independent GP models for each of the four output responses (Horizontal Left, Horizontal Right, Vertical Up, and Vertical Down), treating each output dimension as a scalar response. This approach provides separate calibration parameters $u_k$ ($k=1,2,3,4$) and posterior predictive distributions for each output. The multi-output decision analysis then proceeds by drawing joint realizations across all four dimensions and minimizing the sum of squared responses $G(\mathbf{y}) = \sum_{k=1}^4 y_k^2$.

Following the procedure described in subsection \ref{subsec:param_estimation}, we estimated the calibration parameter $\hat{u}$ and its approximate posterior distribution for each of the four output responses $(y_1, y_2, y_3, y_4)$. Table~\ref{tab:calibration_u} presents the point estimates and the corresponding $95\%$ posterior summary intervals obtained from the LOO samples.

\begin{table}[h]
\centering
\caption{Calibration parameter estimates $\hat{u}$ for the four outputs in Case Study II}
\label{tab:calibration_u}
\begin{tabular}{llll}
\hline
Output & Point Estimate $\hat{u}$ & 95\% Lower Bound & 95\% Upper Bound \\
\hline
$y_1$ & -0.0123 & -0.0323 & 0.0344 \\
$y_2$ & -0.0888 & -0.1154 & -0.0657 \\
$y_3$ & 0.2894 & 0.2010 & 0.3586 \\
$y_4$ & 0.2193 & 0.1614 & 0.2720 \\
\hline
\end{tabular}
\end{table}

These results reveal heterogeneous scaling relationships across the four output dimensions. For $y_1$ (Horizontal Left), the calibration parameter $\hat{u}_1 = -0.0123$ is close to zero with a posterior interval containing zero. This indicates that the low-fidelity simulation lacks a clear linear correlation with the experimental observations for this specific response. For $y_2$ (Horizontal Right), the small negative value $\hat{u}_2 = -0.0888$ suggests that the simulation trends do not align with the experimental data on this wall, which may stem from unmodeled physical complexities or inherent variability in the sparse experimental measurements. In contrast, for the vertical displacements ($y_3$ and $y_4$), the positive parameters ($\hat{u}_3 = 0.2894$ and $\hat{u}_4 = 0.2193$) indicate that the simulation captures the general direction of warpage but systematically overestimates its magnitude, thus requiring down-scaling.

We use the approximate posterior distributions for the calibration parameters to generate $N_u = 100$ posterior samples. Since there are four outputs, each sample represents a vector $\hat{\mathbf{u}}^{(s)} = (\hat{u}_1^{(s)}, \hat{u}_2^{(s)}, \hat{u}_3^{(s)}, \hat{u}_4^{(s)})^\top$. For each sampled vector $\hat{\mathbf{u}}^{(s)}$, we randomly generate $n_{rep} = 100$ sets of $200 \times 4$ Latin hypercube designs (LHDs) as candidate input matrices $\mathbf{X}_{cand}$. Following Algorithm \ref{alg:decision_analysis}, we draw joint realizations for all four output variables from their posterior 
predictive distributions. For each realization, we identify the input setting that minimizes the objective function $G(\mathbf{y}) = \sum_{k=1}^4 y_k^2$.

\begin{figure}[H]
\centering
\includegraphics[width=0.8\textwidth]{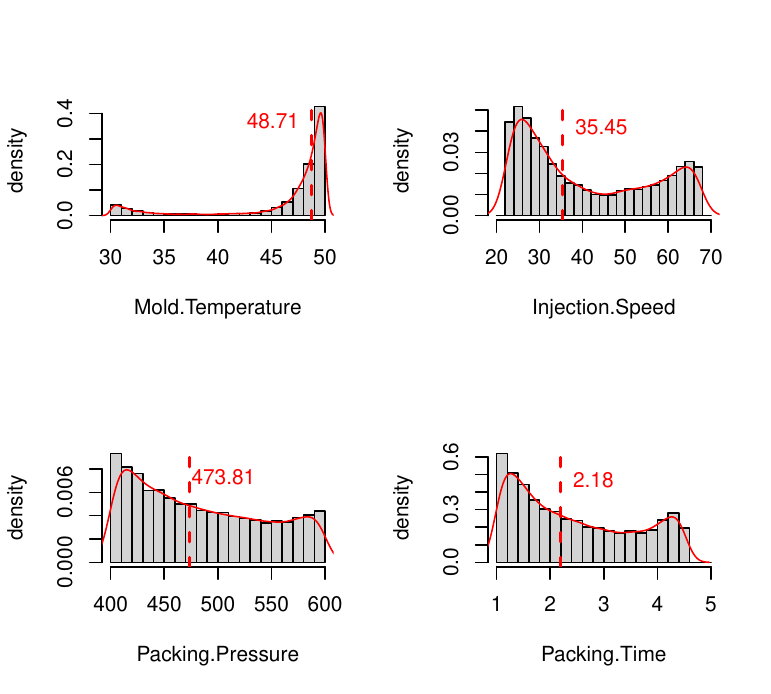}
\caption{Distribution of the optimal candidates in $\mathcal{X}_{opt}$ obtained by applying the proposed framework to the injection molding case study.}
\label{fig:real_minima_distribution}
\end{figure}

The median of the approximate distribution of optimal input locations is approximately $(48.7, 35.5, 473.8, 2.2)$ for (mold temperature, injection speed, packing pressure, packing time), respectively. This represents the most probable optimal process 
parameter setting according to our multi-fidelity Bayesian calibration framework. The distribution of these optimal input locations provides a quantitative assessment of decision uncertainty. Figure \ref{fig:real_minima_distribution} visualizes the variability in optimal decisions across the $N_u \times n_{rep}$ posterior samples.

\section{Conclusion}
\label{sec:conclusion}

In this paper, we developed a decision analysis framework to optimize manufacturing processes. This framework integrates abundant low-fidelity computer simulations with limited high-fidelity experimental data. By using the KOH framework, our method connects efficient models with physical reality and accounts for model discrepancy.

Our main contribution is a five-stage workflow that focuses on the uncertainty of the optimal decision itself. We introduced a simulation-based algorithm to characterize the distribution of optimal input settings. This helps engineers visualize the optimal region and check if the recommended parameters are robust against model errors and noise.

We applied this framework to an illustrative example and two engineering cases: composite cure optimization and injection molding. In the illustrative example, the multi-fidelity approach outperformed single-fidelity methods by correcting simulation bias and reducing uncertainty. For the real-world applications, the results demonstrated that the proposed framework can successfully integrate limited experimental data with simulations to identify stable and robust optimal settings. This shows the practical value of the method in supporting decision-making under real manufacturing constraints.

\section*{Acknowledgments}
Research supported as part of the AIM for Composites, an Energy Frontier Research Center funded by the U.S. Department of Energy (DOE), Office of Science, Basic Energy Sciences (BES), under Award \#DE-SC0023389 (experiments and data analysis). Qiong Zhang acknowledges support by the National Science Foundation (NSF) under Award \#2413630 (statistical methodological development).

\bibliographystyle{chicago}
\bibliography{Reference}

\end{document}